# Multi-microphone Complex Spectral Mapping for Utterance-wise and Continuous Speech Separation

Zhong-Qiu Wang, Peidong Wang, and DeLiang Wang, *Fellow, IEEE*

*Abstract*—We propose multi-microphone complex spectral mapping, a simple way of applying deep learning for time-varying non-linear beamforming, for speaker separation in reverberant conditions. We aim at both speaker separation and dereverberation. Our study first investigates offline utterance-wise speaker separation and then extends to block-online continuous speech separation (CSS). Assuming a fixed array geometry between training and testing, we train deep neural networks (DNN) to predict the real and imaginary (RI) components of target speech at a reference microphone from the RI components of multiple microphones. We then integrate multi-microphone complex spectral mapping with minimum variance distortionless response (MVDR) beamforming and post-filtering to further improve separation, and combine it with frame-level speaker counting for block-online CSS. Although our system is trained on simulated room impulse responses (RIR) based on a fixed number of microphones arranged in a given geometry, it generalizes well to a real array with the same geometry. State-of-the-art separation performance is obtained on the simulated two-talker SMS-WSJ corpus and the real-recorded LibriCSS dataset.

## I. Introduction

DRAMATIC progress has been made in talker-independent speaker separation since deep clustering [1] and permutation invariant training (PIT) [2] were proposed to address the label permutation problem. To improve separation, subsequent studies leverage spatial information afforded by microphone arrays [3]–[6], frequency-domain phase estimation [7], time-domain optimization [8], complex ratio masking [9], and extra information such as speaker embeddings [10] and visual cues [11].

Our study tackles speaker separation in reverberant conditions from the angle of microphone array processing. Since target speakers are directional sources with distinct spatial origins, spatial information provides a potentially important cue for speaker separation. Conventionally, multi-microphone beamforming followed by monaural post-filtering is the most widely adopted approach for multi-channel speech separation [12], [13]. This approach requires the accurate estimates of direction of arrival, power spectral density or spatial covariance matrix. With the recent introduction of deep learning in microphone array processing, all these estimates can now be dramatically improved. The key idea is to use DNN to identify time-frequency (T-F) units dominated by a single source and perform spatial processing based on these T-F units that contain cleaner phase. Representative work includes masking-based beamforming [14]–[16] and speaker localization [17], where DNNs are trained on spectral features to estimate a T-F mask for each microphone, and the estimated masks at different microphones are then pooled to identify T-F units dominated by the same source at all the microphones for covariance matrix computation. Subsequent studies incorporate spatial features such as inter-channel phase differences (IPD) [18], [4], target direction compensated IPD [3], beamforming results [3], and stacked phases and magnitudes [19] as a way of leveraging spatial information to improve mask estimation. However, these studies aim at improving mask or magnitude estimation and do not deal with phase. In addition, they assume that these models are designed for arrays of unknown geometry. Although this generality is desirable, in real-world products such as Amazon Echo and Google Home, the number of microphones and their geometry are fixed. How to leverage fixed-array geometry is a potentially important issue for multi-channel speech processing.

Assuming fixed array geometry, we propose a multi-microphone complex spectral mapping approach for speaker separation, where the real and imaginary (RI) components of multiple microphones are input to a DNN to predict the RI components of the direct-path target speakers captured at a reference microphone. The initial separation results can be utilized to compute target and non-target spatial covariance matrices for MVDR beamforming. The RI components of the beamforming results can be combined with the RI components of multiple microphone signals for post-filtering.

Why would this approach work? We believe that, for a fixed-geometry array, the inter-channel phase patterns are almost the same for signals coming from a particular direction, and the DNN could learn to separate speech arriving from a particular direction by exploiting the fixed and stable spatial information contained in multiple microphones. This approach is in a way similar to classification based sound source localization for

Manuscript received on Oct. 4, 2020; resubmitted on Feb. 6, 2021; revised on Apr. 28, 2021; accepted May 20, 2021. This research was supported in part by an NIDCD grant (R01 DC012048), an NSF grant (ECCS-1808932), and the Ohio Supercomputer Center.

Zhong-Qiu Wang was with the Department of Computer Science and Engineering, The Ohio State University, Columbus, OH 43210-1277 USA, while performing this work. He is now with Mitsubishi Electric Research Laboratories, Cambridge, MA 02139, USA (e-mail: wang.zhongqiu41@gmail.com).

Peidong Wang is with the Department of Computer Science and Engineering, The Ohio State University, Columbus, OH 43210-1277 USA (e-mail: wang.7642@osu.edu).

DeLiang Wang is with the Department of Computer Science and Engineering & the Center for Cognitive and Brain Sciences, The Ohio State University, Columbus, OH 43210-1277 USA (e-mail: dwang@cse.ohio-state.edu).

fixed-geometry arrays [20], where a DNN is trained to learn a one-to-one mapping from inter-channel phase patterns to discretized target directions. Based on deep learning, the proposed approach simultaneously exploits the spectral and spatial information contained in multi-channel inputs to directly predict target speech. The learned DNN model itself can be considered as a non-linear beamformer. In contrast, conventional beamforming techniques compute a filter based on estimated covariance matrices to linearly combine multi-channel signals [12].

A key question is, can a DNN trained using simulated RIRs generated by an RIR simulator based on a given geometry generalize to a real array with the same geometry? An affirmative answer is far from clear, as real recordings exhibit various mismatches from training, such as channel variations and different acoustic environments. In addition, the geometry of a real array, even well calibrated, contains manufacturing imperfections, meaning that the actual geometry would be slightly different from theoretical design. Based on LibriCSS [21], a real-recorded dataset designed for continuous speech separation, we show that our trained models generalize reasonably well to a real array, producing state-of-the-art separation on LibriCSS.

Our study makes three major contributions:

(1) We propose multi-microphone complex spectral mapping, a simple and effective way of using deep learning for time-varying non-linear beamforming on fixed-geometry arrays. Compared with single-microphone complex spectral mapping [22], [23], which trains DNNs based on single-microphone RI components, in multi-microphone complex spectral mapping we directly train DNNs on stacked multi-channel RI components. This simple extension can effectively exploit the spectral and spatial information contained in multiple microphones, producing clear improvements over monaural complex spectral mapping while introducing a negligible number of parameters and a small amount of computation when used with convolutional neural networks. It also shows better performance over a time-invariant MVDR (TI-MVDR) beamformer [12], [22], [23], a strong time-varying MVDR beamformer [24], [23], and a strong time-domain approach [25], [5]. In Section IV.A, we discuss in more details why this supervised learning based approach works well on fixed-geometry arrays;

(2) We integrate multi-microphone complex spectral mapping with conventional MVDR beamforming and post-filtering for better separation. For MVDR beamforming, we design a circular shift mechanism for utilizing a single trained multi-microphone model to compute spatial covariance matrices. For post-filtering, we find that enhancing target speakers one by one rather than predicting them all at once deals with reverberation better. Our experiments show that the immediate separation outputs from our multi-microphone models yield much better recognition results than TI-MVDR, while many previous studies found that TI-MVDR works better for robust ASR [4], [14]–[16], [21], [23], [26], as it produces low speech distortion;

(3) We demonstrate that the trained multi-microphone models based on a simulated array generalize reasonably well to a real device with a matched geometry.

These contributions together lead to state-of-the-art separation performance on the public SMS-WSJ [27] and LibriCSS [21] datasets recently constructed for utterance-wise and continuous-input speaker separation. An earlier version [28] of this study has been published in ICASSP 2020, but it only tackles speech dereverberation, not separation.

The rest of this paper is organized as follows. We present the physical model and objectives in Section II. In Section III, we extend single-microphone complex spectral mapping, which has been successfully applied to geometry-invariant multi-channel speech dereverberation [22] and enhancement [23], to multi-channel speaker separation. In Section IV, we extend this technique to perform multi-microphone complex spectral mapping on fixed-geometry arrays. Section V details the DNN architectures of our models. Experimental setup and evaluation results are detailed in Section VI and VII. Section VIII concludes this paper.

## II. PHYSICAL MODEL AND OBJECTIVES

Given a $P$-channel time-domain signal $\boldsymbol{y}[n] \in \mathbb{R}^{P \times 1}$ recorded in a noisy-reverberant setting with $C$ speakers, the physical model in the short-time Fourier transform (STFT) domain is formulated as

$$\begin{aligned} \boldsymbol{Y}(t,f) &= \sum_{c=1}^{C} \boldsymbol{X}(c,t,f) + \boldsymbol{N}(t,f) \\ &= \sum_{c=1}^{C} \big(\boldsymbol{S}(c,t,f) + \boldsymbol{H}(c,t,f)\big) + \boldsymbol{N}(t,f) \\ &= \sum_{c=1}^{C} \big(\boldsymbol{d}(c,f;q) S_q(c,t,f) + \boldsymbol{H}(c,t,f)\big) + \boldsymbol{N}(t,f), \end{aligned} \quad (1)$$

where $\boldsymbol{Y}(t,f)$, $\boldsymbol{S}(c,t,f)$, $\boldsymbol{X}(c,t,f)$, and $\boldsymbol{N}(t,f) \in \mathbb{C}^{P \times 1}$ respectively denote the complex STFT vectors of the received mixture, direct-path signal of speaker $c$, reverberant image of speaker $c$, and reverberant noise, at time $t$ and frequency $f$. Assuming that each speaker does not move within a single utterance, we have $\boldsymbol{X}(c,t,f) = \boldsymbol{S}(c,t,f) + \boldsymbol{H}(c,t,f) = \boldsymbol{d}(c,f;q) S_q(c,t,f) + \boldsymbol{H}(c,t,f)$, where $S_q(c,t,f) \in \mathbb{C}$ is the complex STFT coefficient of the direct-path signal of source $c$ captured by a reference microphone $q$, $\boldsymbol{d}(c,f;q)$ the time-invariant relative transfer function (RTF) of source $c$ with respect to microphone $q$ and with the $q^{\text{th}}$ element equal to one, and $\boldsymbol{H}(c,t,f)$ the early reflections plus late reverberation of source $c$. In the following sections, when we drop $t$ and $f$ from the notation, we refer to the corresponding complex spectrogram. For example, $S_q(c)$ denotes the spectrogram of speaker $c$ at microphone $q$, and $Y_q$ denotes that of the mixture.

Our goal is to estimate $S_q(c)$ for each source at the reference microphone based on the spectral and spatial information contained in the multi-channel mixture.

Our study assumes a uniform circular array geometry. This type of geometry is very common, including two-microphone linear arrays (or two microphones arranged in a binaural setup), three-microphone equilateral-triangle arrays and four-microphone square arrays. We assume that the same array is used for training and testing. The first microphone on the circle is always considered as the reference microphone, i.e. $q = 1$.

## III. Single-Microphone Complex Spectral Mapping

Figure 2 illustrates one of the proposed systems, SISO$_1$-BF-SISO$_2$, which contains a single-microphone separation network and a single-microphone enhancement network, and a TI-MVDR beamforming module in between. The separation network performs single-microphone complex spectral mapping at each microphone to compute initial separation results, which are then aligned across microphones and used to compute, for each source, a TI-MVDR beamformer to point towards and perform beamforming on that source. The beamforming results are combined with the mixture, i.e. $Y_q$, and the outputs of the first network to train a single-microphone complex spectral mapping based enhancement network to enhance all the target speakers. The MVDR beamforming results are considered as a spatial feature [3], which encodes spatial information and can be used by DNNs to improve separation. As the two networks essentially do single-channel modeling, SISO$_1$-BF-SISO$_2$ can be applied to arrays with diverse geometry. This section describes each module in SISO$_1$-BF-SISO$_2$.

### A. SISO$_1$

We employ single-microphone complex spectral mapping [29], [ ] for both separation and dereverberation. The key idea is to predict the RI components of direct sound from the mixture. We denote this method as **SISO$_1$** (single-microphone input and single-microphone output). See Figure 1 for an illustration. Building upon utterance-level PIT (uPIT) [2], the loss function is defined on the predicted RI components and the resulting magnitude, following [22] and [23],

$$\mathcal{L}_{q,\text{uPIT}} = \min_{\psi_q \in \Psi} \sum_{c=1}^{C} \Big( \left\| \hat{R}_q^{(k)}(\psi_q(c)) - \text{Real}\big(S_q(c)\big) \right\|_1 \\ + \left\| \hat{I}_q^{(k)}(\psi_q(c)) - \text{Imag}\big(S_q(c)\big) \right\|_1 \\ + \left\| \sqrt{\hat{R}_q^{(k)}(\psi_q(c))^2 + \hat{I}_q^{(k)}(\psi_q(c))^2} - |S_q(c)| \right\|_1 \Big), \quad (2)$$

where $\Psi$ denotes the set of all the permutations of $C$ sources, $\psi_q$ refers to a permutation (or a pairing of speaker and DNN output) at microphone $q$. $\hat{R}_q$ and $\hat{I}_q$ are the estimated RI components produced by linear activation in the output layer, Real($\cdot$) and Imag($\cdot$) respectively extract the real and imaginary components, $|\cdot|$ computes magnitude, $\|\cdot\|_1$ computes the $L_1$ norm, and $k \in \{1,2\}$ denotes which DNN produces the output since we will have two DNNs in our later multi-channel system. The separation result is obtained as $\hat{S}_q^{(k)} = \hat{R}_q^{(k)} + j\hat{I}_q^{(k)}$. The network input is the RI components of $Y_q$. We will describe the DNN architecture in Section V.

Following uPIT, we assume that there are at most $C$ speakers in each mixture in the offline utterance-wise case and at most $C$ speakers in each block in the block-online continuous case, for both system training and deployment.

### B. SISO$_1$-BF and MVDR Beamforming

For multi-channel processing, we apply SISO$_1$ to each microphone and use the predicted multi-channel complex spectra to compute statistics for MVDR beamforming (denoted as **SISO$_1$-BF**). See the TI-MVDR part in Figure 2.

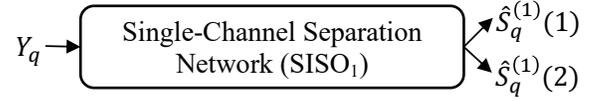

Figure 1. SISO$_1$ system for 2-speaker separation and dereverberation.

We emphasize that before beamforming, the source alignment module in Figure 2 is needed to align uPIT results across microphones, as SISO$_1$ is applied to each microphone independently and the uPIT results at different microphones may exhibit different permutations of speakers. The prime in, say, $\hat{S}_q^{(1)\prime}$ in Figure 2 is used to differentiate the notation from $\hat{S}_q^{(1)}$, as we align speakers across microphones. The alignment is done by simply aligning the outputs at each non-reference microphone to the outputs at the reference microphone based on their magnitude distance.

We use estimated complex spectra to compute target and non-target covariance matrices, $\hat{\boldsymbol{\Phi}}^{(s)}(c,f)$ and $\hat{\boldsymbol{\Phi}}^{(v)}(c,f)$, for MVDR beamforming

$$\hat{\boldsymbol{\Phi}}^{(s)}(c,f) = \frac{1}{T} \sum_t \hat{\boldsymbol{S}}(c,t,f) \hat{\boldsymbol{S}}(c,t,f)^\text{H} \\ \hat{\boldsymbol{\Phi}}^{(v)}(c,f) = \frac{1}{T} \sum_t \hat{\boldsymbol{V}}(c,t,f) \hat{\boldsymbol{V}}(c,t,f)^\text{H}, \quad (3)$$

where $T$ is the total number of frames within a mixture (or a sliding block) and $\hat{\boldsymbol{V}}(c,t,f) = \boldsymbol{Y}(t,f) - \hat{\boldsymbol{S}}(c,t,f)$. An earlier way [14], [3] of using DNNs to compute covariance matrices is by first using a DNN to estimate a real-valued mask $\lambda(c,t,f)$ for each source at each T-F unit, and then using it to compute a weighted sum of mixture outer products, namely $\frac{1}{T} \sum_t \lambda(c,t,f) \boldsymbol{Y}(t,f) \boldsymbol{Y}(t,f)^\text{H}$. It is suggested in [22], [23], [30] that Eq. (3) leads to better covariance matrix estimation than using real-valued masks, as long as the estimated complex spectra exhibit better phase than the mixture.

As a target speaker can be viewed as a point source, following [15], [12] we compute its steering vector $\hat{\boldsymbol{r}}(c,f)$ as follows

$$\hat{\boldsymbol{r}}(c,f) = \mathcal{P}\{\hat{\boldsymbol{\Phi}}^{(s)}(c,f)\} \quad (4)$$
$$\hat{\boldsymbol{d}}(c,f;q) = \hat{\boldsymbol{r}}(c,f)/\hat{r}_q(c,f), \quad (5)$$

where $\mathcal{P}\{\cdot\}$ extracts the principal eigenvector. We further divide $\hat{\boldsymbol{r}}(c,f)$ by its $q^\text{th}$ element to obtain an estimate of the RTF with respect to the reference microphone.

An MVDR beamformer is computed as

$$\hat{\boldsymbol{w}}(c,f;q) = \frac{\hat{\boldsymbol{\Phi}}^{(v)}(c,f)^{-1} \hat{\boldsymbol{d}}(c,f;q)}{\hat{\boldsymbol{d}}(c,f;q)^\text{H} \hat{\boldsymbol{\Phi}}^{(v)}(c,f)^{-1} \hat{\boldsymbol{d}}(c,f;q)} \quad (6)$$

and beamforming results $\widehat{BF}_q(c,t,f)$ are computed as

$$\widehat{BF}_q(c,t,f) = \hat{\boldsymbol{w}}(c,f;q)^\text{H} \boldsymbol{Y}(t,f). \quad (7)$$

### C. SISO$_1$-BF-SISO$_2$

Next, the MVDR beamforming results are combined with the mixture and the SISO$_1$ separation results, all at the reference microphone, to train another SISO network to improve the separation (see Figure 2). This second SISO network is essentially a post-filter, which performs enhancement and does not need to resolve the permutation problem. It estimates all the $C$ speakers by using $\langle Y_q, \widehat{BF}_q(1), \ldots, \widehat{BF}_q(C), \hat{S}_q^{(1)}(1), \ldots, \hat{S}_q^{(1)}(C) \rangle$ as inputs to predict $\langle S_q(1), \ldots, S_q(C) \rangle$ (denoted as **SISO$_1$-BF-SISO$_2$**). The loss function is defined on the predicted RI components and their magnitudes, similar to Eq. (2) but without resolving permutations. Note that we use different subscripts, say SISO$_1$-BF-SISO$_2$, to denote different SISO models, as they take in different features. This convention applies to all of our models.

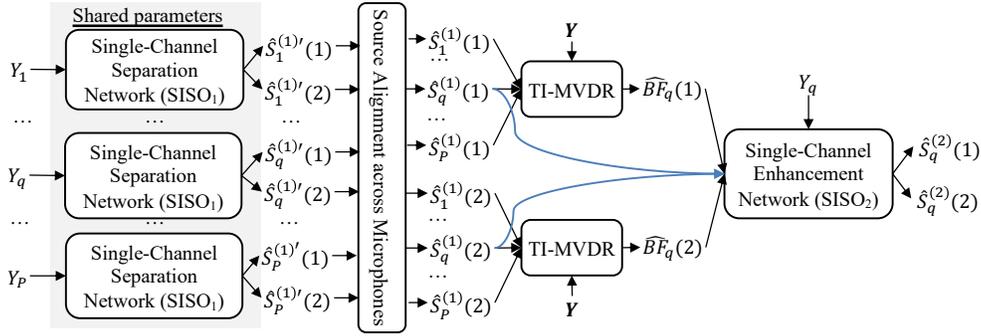

Figure 2. SISO$_1$-BF-SISO$_2$ system for two-speaker separation and dereverberation.

Although the second network takes in the MVDR beamforming results in its inputs, we still consider it doing single-channel complex spectral mapping, where the beamforming results are viewed as a spatial feature [3] that can leverage spatial information to improve separation.

We tried to replace the TI-MVDR in SISO$_1$-BF-SISO$_2$ with various time-varying beamformers [23], [31]. However, we do not observe clearly better performance, likely because the sources do not move within each mixture in this study.

### IV. MULTI-MICROPHONE COMPLEX SPECTRAL MAPPING

For fixed-geometry arrays, we replace each SISO network in the SISO$_1$, SISO$_1$-BF and SISO$_1$-BF-SISO$_2$ systems with a MISO (multi-microphone input and single-microphone output) network, which includes multi-microphone inputs as features for multi-microphone complex spectral mapping. This leads to our MISO$_1$, MISO$_1$-BF and MISO$_1$-BF-MISO$_2$ systems (see Figure 3(a) and (b)). Each one of them is better than its single-microphone counterpart, since the DNNs are trained directly on multi-microphone inputs to leverage spatial information. The following subsections describe each of the systems. Section IV.C introduces a MISO$_1$-BF-MISO$_3$ system (see Figure 3(c)), where the second DNN enhances target speakers one by one, rather than enhancing all of them at once. The last subsection discusses the application of MISO$_1$-BF-MISO$_3$ with a speaker counting module for block-online CSS.

### A. MISO$_1$

In a multi-microphone setup, SISO$_1$-BF-SISO$_2$ does not use DNNs to directly model multiple microphones. We propose **MISO$_1$** networks, illustrated in Figure 3(a), for multi-channel speaker separation, where we stack the RI components of multiple microphone signals $\langle Y_q, \ldots, Y_P, Y_1, \ldots, Y_{q-1} \rangle$ to predict the RI components of all the speakers $\langle S_q(1), \ldots, S_q(C) \rangle$ at a reference microphone $q$. The loss function is $\mathcal{L}_{q,\text{uPIT}}$. We will talk about the DNN architecture in Section V.

This approach is in spirit similar to the classic multi-channel Wiener filter [12], where a linear filter is computed per T-F unit to project the mixture onto target speech. Our study trains a DNN to do this. Implicitly broadband and capable of exploiting a large context window along time and frequency by using, for example, dilated convolution, recurrence or self-attention, the DNN essentially learns to perform time-varying non-linear beamforming. Although this would be difficult to learn for unknown arrays, where test geometry differs from the trained geometry, it would likely work well if array geometry is the same between training and testing, as the inter-channel phase patterns do not change for signals coming from a particular direction. In such a case, the DNN would likely be able to exploit such fixed patterns for better separation than using just a single microphone. This approach is conceptually simple, computationally efficient, and can be easily modified for online real-time processing.

Different from the convolutional beamformer approach [32] and approaches that use DNN to first predict beamforming filters and then perform linear filtering [33], [5], [34], our approach directly uses a DNN to predict target speech from multi-channel inputs and the DNN itself is the beamformer. Another related study [19] stacks the magnitude and phase of the multi-channel mixture as inputs to a DNN to estimate a real-valued mask, which can be used for direct enhancement or computing time-varying speech and noise covariance matrices for beamforming. However, this approach does not use DNN for phase estimation. In addition, the noise covariance matrix is computed based on recursive averaging, which usually cannot lead to sufficient noise suppression at each T-F unit, because the covariance matrix computed based on averaging more frames surrounding a T-F unit would be more different from the instantaneous noise outer product that can, in the oracle case, lead to perfect noise suppression at that T-F unit.

Although there are time-domain approaches using multi-microphone waveforms as the inputs for DNNs to predict target waveforms at a reference microphone for speech enhancement and speaker separation, similar to the proposed MISO approach [5], [6], [25], [35], [36], their success in environments with significant reverberation is less impressive [37] than in anechoic conditions, and their generalization to realistic noisy-reverberant recordings is unclear. In addition, our study integrates multi-microphone complex spectral mapping with beamforming and post-filtering, which produce further improvements.

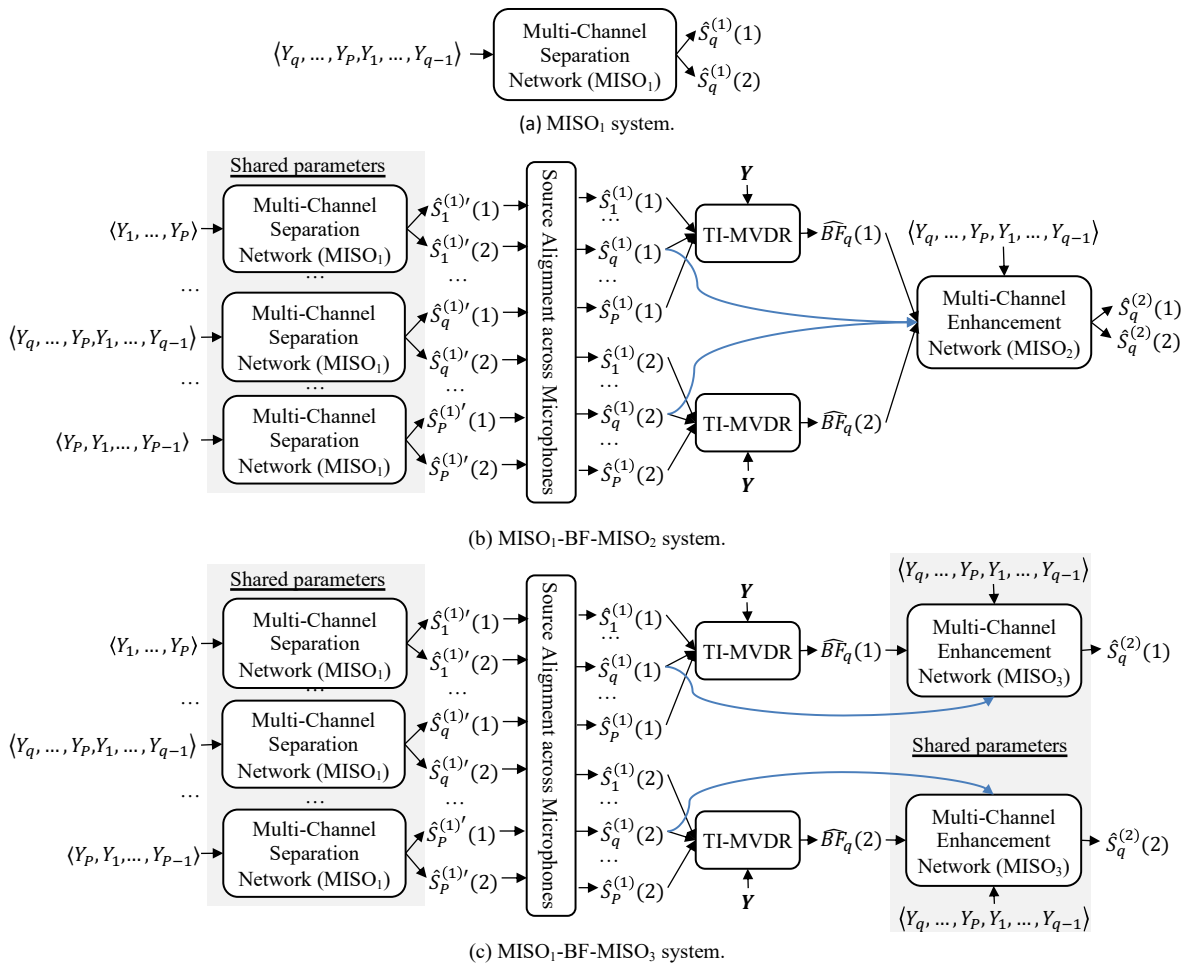

Figure 3. (a) MISO$_1$; (b) MISO$_1$-BF-MISO$_2$; and (c) MISO$_1$-BF-MISO$_3$ systems for two-speaker separation and dereverberation.

### B. MISO$_1$-BF

Similar to SISO$_1$-BF, we use the separation results by MISO$_1$ to compute an MVDR beamformer for each source (denoted as **MISO$_1$-BF**). See the TI-MVDR part in Figure 3(b). Since MISO is trained on multi-channel inputs, it can provide better signal statistics for MVDR than SISO.

To compute covariance matrices using Eq. (3), we need to have an estimate of the target speech at each microphone. Since in our experiments MISO$_1$ is trained on the concatenation of an ordered list of microphones $\langle Y_1, \dots, Y_P \rangle$ to predict $S_1$, at run time we cannot feed in $\langle Y_1, \dots, Y_P \rangle$ to MISO$_1$ to estimate say $S_2$. One cumbersome way is to train another model to predict $S_2$. In this way, one has to train $P$ different models, one at each microphone. We instead circularly shift the microphones at run time for the prediction of each microphone signal, i.e. we feed $\langle Y_p, \dots, Y_P, Y_1, \dots, Y_{p-1} \rangle$ to MISO$_1$ to predict $S_p$ for $p \in \{1, \dots, P\}$, essentially rotating the array. This strategy should work if the microphones are arranged uniformly on a circle, since using $\langle Y_p, \dots, Y_P, Y_1, \dots, Y_{p-1} \rangle$ as inputs to predict $S_p$ is essentially the same as what MISO$_1$ is trained to do.

What if microphones are configured in a popular Amazon Echo setup where the first $P-1$ microphones are on a circle and the last at the circle center? In such a case, we can circularly shift the microphones on the circle, and always put the center microphone at last in the ordered list, i.e. we use $\langle Y_p, \dots, Y_{P-1}, Y_1, \dots, Y_{p-1}, Y_P \rangle$ to predict $S_p$, for $p \in \{1, \dots, P-1\}$. In this case, we can only use the $P-1$ microphones on the circle for later MVDR beamforming, which should not be much worse than using $P$ microphones if $P$ is not small and the aperture sizes are the same. An alternative is to use T-F masks estimated on the $P-1$ microphones to compute a pooled mask to perform mask-based beamforming on $P$ microphones. When $P$ is large, the quality of the pooled mask computed from $P-1$ masks should be very close to that computed from $P$ masks. We leave this alternative for future investigation.

We cannot use this shifting trick for many non-circular arrays such as linear arrays with more than two microphones. In such cases, we can train a multi-microphone input and multi-microphone output (MIMO) network that predicts the target speech at all the microphones [28]. However, in the circular-array case, a MIMO network is found to produce worse estimation of target speech at each microphone than a MISO network, because MIMO has more signals to predict than MISO [28].

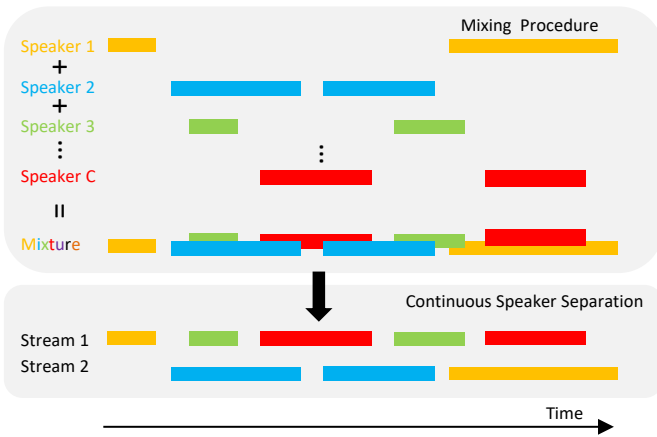

Figure 4. Task illustration of continuous speech separation.

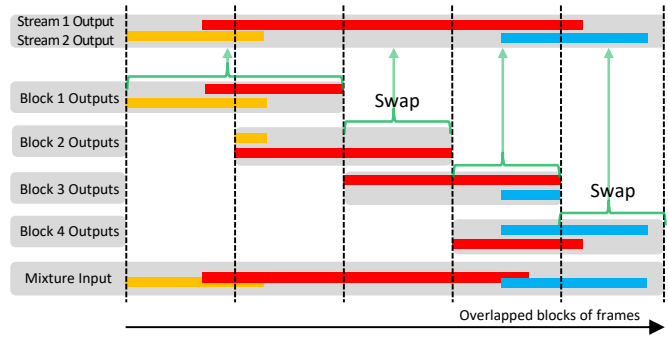

Figure 5. Illustration of block-online CSS. The separation results in the overlapped frames between consecutive blocks are used for block stitching. In block 2 and 4, the two estimates are swapped in the stitching process.

### C. MISO$_1$-BF-MISO$_2$ and MISO$_1$-BF-MISO$_3$

Different from SISO$_1$-BF-SISO$_2$, the beamforming results $\widehat{BF}_q$ are combined with the multi-channel inputs and initial separation results $\hat{S}_q^{(1)}$ to train another MISO network to further predict $S_q$ (see Figure 3(b) or (c)). This MISO network is designed to leverage multi-microphone modeling for post-filtering. It essentially performs enhancement and does not need to resolve the permutation problem, as the problem has already been resolved by the first network. It can estimate $C$ speakers all at once by using $\langle Y_q, \ldots, Y_P, Y_1, \ldots, Y_{q-1}, \widehat{BF}_q(1), \ldots, \widehat{BF}_q(C), \hat{S}_q^{(1)}(1), \ldots, \hat{S}_q^{(1)}(C)\rangle$ as inputs to predict $\langle S_q(1), \ldots, S_q(C)\rangle$ (denoted as **MISO$_1$-BF-MISO$_2$**, see Figure 3(b)), or predict each target speaker one by one by using $\langle Y_q, \ldots, Y_P, Y_1, \ldots, Y_{q-1}, \widehat{BF}_q(c), \hat{S}_q^{(1)}(c)\rangle$ to predict $S_q(c)$ (denoted as **MISO$_1$-BF-MISO$_3$**, see Figure 3(c)). We find that the latter produces better separation, likely because each speaker is convolved with a different RIR and hence it is better to enhance the speakers individually. In addition, the DNN only needs to model the pattern of a single target speaker rather than that of multiple speakers combined, and does not need to learn to deal with the varying energy levels between the speakers. The downside is that the second DNN needs to be used $C$ times at run time, once for each speaker.

Although the MISO networks in MISO$_1$-BF-MISO$_3$ (or MISO$_1$-BF-MISO$_2$) can be viewed as non-linear beamformers, we find that adding the TI-MVDR beamformer in between dramatically improves the performance of the second network. This is likely because an MVDR beamformer is built based on signal processing principles. It can produce reliable separation especially in conditions with low reverberation and noise, if it is pointed towards a target speaker and puts null beams on the other speakers. Such reliable separation could provide complementary information to boost MISO based separation. In contrast, the MISO models alone are built from supervised learning. They are data-driven and cannot leverage the gains provided by conventional signal processing. To show the benefits of including a TI-MVDR beamformer, in our experiments we will compare MISO$_1$-BF-MISO$_2$ with a **MISO$_1$-MISO$_4$** baseline, where MISO$_4$ is trained in the same way as MISO$_2$, but not taking in the TI-MVDR results as inputs. Similarly, we will also compare MISO$_1$-BF-MISO$_3$ with a **MISO$_1$-MISO$_5$** baseline, where MISO$_5$ is trained in the same way as MISO$_3$, but not taking in the TI-MVDR results.

### D. Block-online MISO$_1$-BF-MISO$_3$ for CSS

Continuous-input speech separation [21] deals with the scenario where signals from an unknown number of speakers, possibly degraded by reverberation, noise and various degrees of speaker overlap, come as a continuous stream. Following [21], we focus on separating the input stream into two streams, each being enhanced and free of concurrent speech, as illustrated in Figure 4. This processing can be useful for streaming for example conversational speech recognition systems.

We follow the overlap-block idea proposed in [21] for block-online CSS. Following [21], we assume that each short block (in our study 2.424 s) contains at most two speakers so that a two-speaker PIT model can be used in each block. This is a reasonable assumption in meeting scenarios, as long as turn taking does not happen frequently. The MISO-BF-MISO model is applied to each block independently, i.e. considering each block as a different mixture and not using any information from future frames. Since our networks produce two estimates at each block, we need to align the two estimates in the current block with those in the previous block so that any continuous speaker segment spanning the two blocks can be put into the same output stream (see Figure 5 for an illustration). This alignment procedure, often referred to as speaker tracking or block stitching [21], is performed by comparing the separation results in the overlapped regions between consecutive blocks. The inevitable delay is the block shift size if a non-causal model is applied in each block. Figure 5 illustrates this idea.

On the LibriCSS dataset used in our experiments, we empirically observe that, when a model trained on two-speaker mixtures is applied to process utterances containing only one speaker, it sometimes cannot put the speaker in one stream and set the other to silence, resulting in very weak but intelligible speech residual in the other stream. To suppress the residual, we train a frame-level speaker counting network to (1) count the number of speakers at each frame of the current block; (2) find segments of frames containing only one speaker based on the frame-level counting results; (3) merge the stream with weaker energy to the other stream for each detected one-speaker segment within the block; and (4) suppress the weaker-stream segment by multiplying it with a small constant. We perform three-class classification (i.e. zero, one or two speakers) for frame-wise speaker counting. The network architecture will be discussed in the following section.

## V. DNN Architecture

Figure 6 shows the DNN architecture of our SISO and MISO models. Similar architectures have shown strong performance in a number of tasks including speaker separation [9], speech dereverberation [22], [28] and speech enhancement [23]. The architecture is a temporal convolutional network (TCN) [38] clamped by a U-Net [39] which includes an encoder for down-sampling and a decoder for up-sampling along frequency. We add DenseNet blocks [40] at multiple frequency scales in the encoder and decoder. The motivation of this network design is that U-Net can maintain local fine-grained structure via its skip connections and model contextual information along frequency through down- and up-sampling, TCN can leverage long-range information by using dilated convolutions along time, and DenseNet blocks encourage feature re-use and improve discriminability. The encoder contains one two-dimensional (2D) convolution, and seven convolutional blocks, each with 2D convolution, exponential linear units (ELU) non-linearity and instance normalization (IN), for down-sampling. The decoder includes seven blocks of 2D deconvolution, ELU and IN, and one 2D deconvolution, for up-sampling. The TCN contains two layers, each with seven dilated convolutional blocks. We use one-dimensional (1D) depth-wise separable convolution in the dilated convolutional blocks to reduce the number of parameters. We stack RI components as features maps in the network input and output. Different models share the same network architecture and differ only in the network input and output. Each SISO or MISO network contains around 6.9 million parameters.

This convolutional encoder-decoder achitecture performs convolution directly on multi-microphone RI components to simultaneously exploit the spectral and spatial information contained in multi-microphone inputs. Increasing the number of microphones only incurs a small number of parameters and a small amount of computation. For the network architecture in Figure 6, a MISO network only has $((P-1) \times 2) \times 24 \times 3 \times 3$ more parameters than a SISO network, where $P$ is the number of microphones, 24 the number of feature maps in the first 2D convolution, $3 \times 3$ the kernel size, and 2 is because we stack real and imaginary components. The increased amount of computation of MISO over SISO is only from the first convolutional layer, which is negligible relative to the rest of the network. In contrast, earlier studies encode spatial information using inter-channel phase patterns [18], [4], [6], [21], [41] by decoupling multi-channel RI components into separate IPDs and magnitudes. As the decoupled features exhibit different patterns less suitable for direct convolution, typical methods [4], [6] use a fully-connected input layer (or 1D convolution) to compress disparate IPDs and magnitudes into lower-dimensional fixed-length vectors, which however may lose phase information before later processing. In addition, these methods introduce many more parameters when the number of microphones increases, because of the fully-connected layer. Further, they typically only consider the IPDs between a reference microphone and other microphones to reduce parameters. This may not be optimal as not all the microphone pairs are leveraged. In contrast, our MISO networks with encoder-decoder structure perform convolution directly on multi-microphone RI components to exploit inter-channel phase patterns among all the microphones. The short-cut connections from the encoder to the decoder (shown as the "Concatenate" lines in Figure 6) can better flow the multi-channel phase information inside the network.

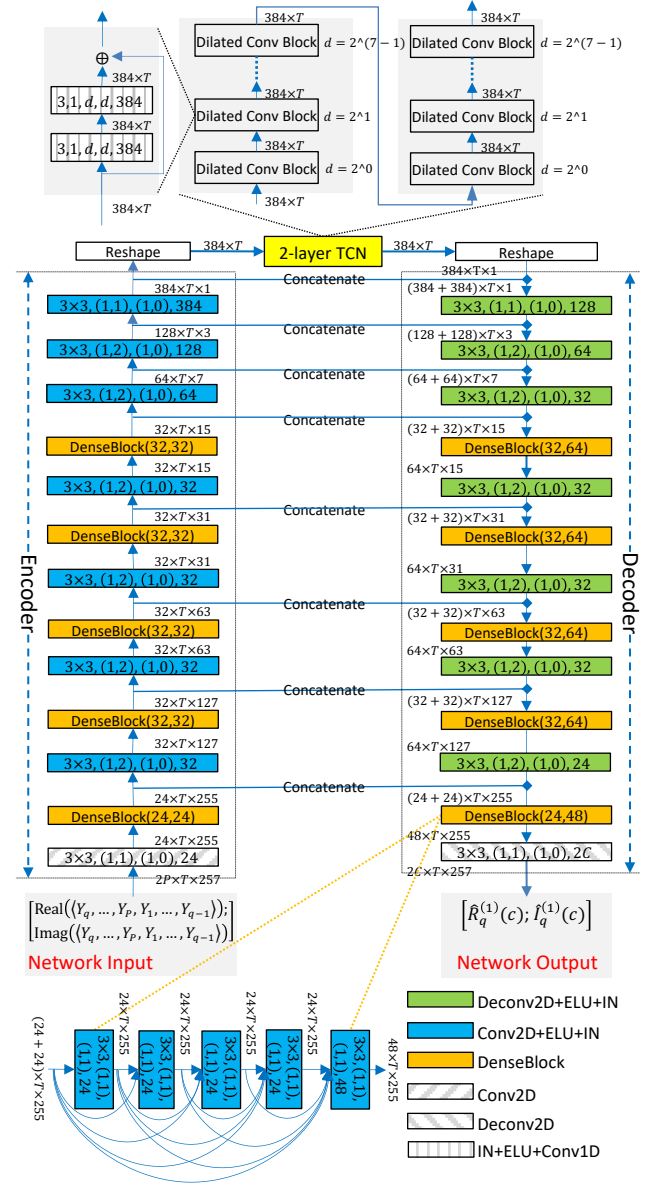

Figure 6. Example network architecture of MISO$_1$ for predicting the RI components of $S_q$ from multi-channel inputs $\langle Y_q, ..., Y_P, Y_1, ..., Y_{q-1} \rangle$. The tensor shape after each encoder-decoder block is in the format: *featureMaps×timeSteps×frequencyChannels*. Each one of Conv2D, Deconv2D, Conv2D+ELU+IN and Deconv2D+ELU+IN blocks is specified in the format: *kernelSizeTime × kernelSizeFreq, (stridesTime, stridesFreq), (paddingsTime, paddingsFreq), featureMaps*. Each DenseBlock($g_1, g_2$) contains five Conv2D+ELU+IN blocks with growth rate $g_1$ for the first four layers and $g_2$ for the last layer. The tensor shape after each TCN block is in the format: *featureMaps × timeSteps*. Each IN+ELU+Conv1D block is specified in the format: *kernelSizeTime, stridesTime, paddingsTime, dilationTime, featureMaps*.

For the speaker counting network used in our CSS algorithm, we only use the encoder and the TCN components, and remove the decoder. We add a softmax layer on the ouptut of TCN for frame-wise classification, and train the model using cross-entropy. The input feature is the same as that in $MISO_1$ in the multi-channel case and $SISO_1$ in the monaural case. On our simulated reverberant two-speaker validation set (introduced later in Section VI.C), the accuracy of our frame-level speaker counting model in the seven-channel case is around 97%, which is quite accurate.

## VI. EXPERIMENTAL SETUP

Our algorithms are evaluated on two datasets, SMS-WSJ [27] and LibriCSS [21]. The first one is for two-talker separation in simulated and matched reverberant conditions, and the second for CSS in real-recorded and unmatched reverberant conditions. Both datasets contain weak environmental noise. This section describes each dataset, our simulated training data for LibriCSS, the training procedure of our DNN models, system configurations for the evaluation on LibriCSS, benchmark systems, and evaluation metrics.

### A. SMS-WSJ Dataset

SMS-WSJ [27] contains two-speaker mixtures in reverberant conditions. The sampling rate is 8 kHz. The clean sources are drawn from the WSJ0 and WSJ1 datasets. The database contains 33561, 982, and 1332 two-speaker mixtures for training, validation, and testing, respectively. The array is circular with six microphones uniformly spaced on a circle with 10 cm radius. The speaker-to-array distance is sampled from the range [1.0, 2.0] m, and the reverberation time (T60) is drawn from the range [0.2, 0.5] s. A weak white noise is added to simulate sensor noises. The speaker angles in each mixture are randomly sampled from $[-\pi, +\pi]$. For ASR, we use the default backend acoustic model, which is trained by using single-speaker reverberant-noisy speech as inputs and the clean alignments of its corresponding direct sound as labels. A default task-standard trigram language model is used for decoding. For separation, we consider direct sound as the training target and perform both dereverberation and separation. This is different from the official SMS-WSJ setup, which considers direct sound plus early reflections as the target for metric computation. We think that this modification is reasonable for ASR tasks, as early reflections smear spectral patterns, although not as severely as late reverberation. It also aligns with beamforming, as beamforming methods are designed for extracting point sources.

In addition to one- and six-channel separation, we consider two-channel separation, based on the first and the fourth microphone on the circle, and three-channel separation, using the first, third and fifth microphones.

### B. LibriCSS Dataset

LibriCSS [21] contains ten hours of conversational speech data recorded by playing LibriSpeech signals through loud speakers in reverberant rooms. The sampling rate is 16 kHz. The task is to perform conversational speech recognition with room reverberation and a wide range of speaker overlaps. There are ten one-hour sessions, each consisting of six ten-minute mini-sessions with different speaker overlap ratios ranging from 0% to 40%, including 0S (no overlap with short inter-utterance silence between 0.1 and 0.5 seconds), 0L (no overlap with long inter-utterance silence between 2.9 and 3.0 seconds), and 10%, 20%, 30% and 40% overlaps. The recording device has seven microphones, with six of them uniformly arranged on a circle with a 4.25 cm radius, and one at the circle center. The distance between loud speakers to the array ranges from 33 to 409 cm. This dataset contains two kinds of ASR evaluations, utterance-wise evaluation and continuous-input evaluation, both expecting frontend processing to produce two streams. The former assumes that each utterance has been accurately segmented and the goal is to recognize the segmented utterance. The ASR backend scores both streams and the one with lower WER is considered the final WER. The latter segments each mini-session to 60- to 120-second long segments, each with 8 to 10 utterances from at most eight speakers. The goal is to recognize all the utterances in each segment. The ASR backend scores both streams, but combines the two decoding results to compute the final WER. This is different from the utterance-wise setup, where the lower WER is picked. As a result, in one-speaker segments the continuous setup requires one stream to contain the speaker and the other to be completely silent. This is why we introduced a speaker counting network in Section IV.D to suppress speech residuals.

### C. Simulated Training Data for LibriCSS

Since LibriCSS only contains testing data, we need to simulate training and validation data for separation by ourselves. Our training data includes 76,750 (~129 hours) seven-channel two-speaker mixtures with moderate levels of room reverberation and weak air conditioning noise. Among all the frames, 12% contain no speaker, 55% one speaker and 33% two speakers. We sample clean source signals from the *train-clean-*{100,360} set of LibriSpeech. Assuming the array geometry of the LibriCSS recording device, we use an RIR generator [42] to simulate seven-microphone RIRs. T60 is sampled from the range [0.2,0.6] s. The average distance between speaker and array center is sampled from [0.75,2.5] m. The average direct-to-reverberation energy ratio of the RIRs is −0.3 dB with 3.9 dB standard deviation. The angles of the two speakers are randomly sampled from $[-\pi, +\pi]$ and ensured to be at least 10° apart. The energy level between the two speakers is sampled from the range $[-7,7]$ dB. We sample an air conditioning noise from the REVERB corpus for each reverberant two-talker mixture. The SNR between the anechoic two-source mixture and the noise is drawn from the range [10,30] dB.

The labels used for training the speaker counting model are obtained by first applying a pre-trained DNN based voice activity detector [43] to the spatialized anechoic signal of each one of the two speakers at the reference microphone, and then combining the two VAD results to get the number of speakers at each frame.

We cut each training mixture into 300-frame segments to train our models. Based on the oracle frame-level VAD results computed from the anechoic speech of each speaker, 24% of these segments have only one speaker and 76% two speakers. One-speaker segments are hence reasonably represented in our training data.

### D. Miscellaneous Configurations for LibriCSS

For offline speaker separation, we normalize the sample variance of each multi-channel signal to one before any processing. This can deal with random gains in mixtures, and would be important for mapping based methods [22], [23]. For block-online processing, we compute sample variance online, i.e. using all the samples up to the current block to normalize the current block. After obtaining the separation results at the current block, we reverse the normalization to recover the original levels before stitching. We normalize input features globally to zero mean and unit variance. When performing global normalization on RI components, the mean is set to zero due to the randomness of phase, and the statistics for standard deviation is collected from both the real and imaginary components within each freqeuncy so that the phase remains the same after scaling RI components. Note that scaling RI components using different factors modifies the underlying phase.

For STFT, the window size is 32 ms, the shift is 8 ms, and the analysis window is the square root of Hann window. We use 512-point discrete Fourier transform (DFT) to extract 257-diminensional complex spectra for 16 kHz sampling rate, and 256-point DFT to extract 129-dimenisonal complex spectra for 8 kHz sampling rate.

Following [21], [41], we set the run-time block size to 2.424 seconds for CSS. It corresponds to 300 frames as our STFT window size is 32 ms and window shift is 8 ms. The block shift is set to 1.2 seconds. This means that our block-online system has a 1.2-second inherent delay, as in [21], [41]. For the utterance-wise evaluation, at run time we use this overlapped-block idea for the first network, as the speaker in a segmented utterance could overlap with the preceding and the succeeding speakers, while we use full-utterance information for beamforming and post-filtering.

As our study focuses on separation, we mainly use the default ASR backend provided by LibriCSS for recognition to facilitate comparisons with or by other studies. We also employ a more powerful end-to-end ASR backend to improve recognition. We feed resynthesized signals to backends for recognition.

### E. Training Procedure

We use our simulated training set (for LibriCSS) and MISO$_1$-BF-MISO$_3$ as an example to describe how we train the systems. MISO$_1$ is trained to predict $S_1$ based on $\langle Y_1, ..., Y_P \rangle$ using the 76,750 seven-channel mixtures. We train the model using 300-frame segments. We then run MISO$_1$ on each full-length training mixture to compute $\widehat{S}(c)$ following the circular shifting idea, and use all the frames in the mixture to compute the covariance matrices for TI-MVDR beamforming (with the reference microphone index set to one). We then train the MISO$_3$ network also on 300-frame segments. The two networks are trained sequentially. All the other systems are trained in a similar way.

All DNNs are trained using the Adam optimizer for at most 100 epochs. The learning rate starts at 0.001 and is halved if the validation loss is not improved for three epochs. We stop the training process when the learning rate decays to 3.125e-5.

### F. Benchmark Systems

Based on SMS-WSJ, we consider the following benchmarks:

(1) We compare MISO$_1$, MISO$_1$-BF and MISO$_1$-BF-MISO$_2$ respectively with SISO$_1$, SISO$_1$-BF and SISO$_1$-BF-SISO$_2$. This can show the effectiveness of direct multi-microphone modeling on fixed-geometry arrays over single-microphone modeling;

(2) We compare MISO$_1$ with a strong DNN-supported time-varying MVDR beamformer, computed by using the outputs of MISO$_1$ to calculate time-varying covariance matrices. This beamformer is considered as an improved version of the beamformer proposed in [19]. This comparison can show the effectiveness of MISO over more conventional time-varying beamforming. Following [24] and [23], the beamformer is computed by replacing the time-invariant non-target covariance matrix $\widehat{\boldsymbol{\Phi}}^{(v)}(c, f)$ in Eq. (6) with a time-varying one

$$\widehat{\boldsymbol{\Phi}}^{(v)}(c,t,f) = \alpha \frac{\sum_{t'=t-\Delta}^{t+\Delta} \widehat{V}(c,t',f)\widehat{V}(c,t',f)^H}{trace\left(\sum_{t'=t-\Delta}^{t+\Delta} \widehat{V}(c,t',f)\widehat{V}(c,t',f)^H\right)/P} \\ + (1-\alpha) \frac{\widehat{\boldsymbol{\Phi}}^{(v)}(c,f)}{trace\left(\widehat{\boldsymbol{\Phi}}^{(v)}(c,f)\right)/P}, \quad (8)$$

where $\Delta$ denotes the context window size in frames on each side and $\alpha$ (set to 0.5 in this study) a weighting term. It is a combination of the short- and long-term estimates of the non-target covariance matrix. The RTF is still computed in a time-invariant way using Eq. (5), as the target speaker is assumed not moving within an utterance. Decreasing $\Delta$ makes $\widehat{\boldsymbol{\Phi}}^{(v)}(c,t,f)$ more time-varying, but also suffer more from the errors in $\widehat{V}(c,t,f)$, as DNNs cannot estimate complex spectra perfectly;

(3) We compare our system with a popular spatial clustering technique provided with SMS-WSJ, which is based on complex angular central GMM (cACGMM) with or without further TI-MVDR beamforming [27]. We consider it as a representative conventional approach for multi-channel speaker separation;

(4) We compare our systems with representative time-domain end-to-end approaches such as the monaural DPRNN-TasNet [44], an improved version of Conv-TasNet [8], the multi-channel FaSNet with TAC modules [5], a representative time-domain beamforming technique extending monaural DPRNN-TasNet for multi-channel separation, and a multi-channel Conv-TasNet [45]. All of them are popular in speaker separation. We implement them using the Asteroid toolkit;

(5) For LibriCSS, we compare our system with two strong block-online systems recently reported in [21] and [41], which use real-valued masking in the monaural case and MVDR beamforming in the multi-channel case.They are proposed by the authors of LibriCSS.

### G. Evaluation Metrics

We consider scale-invariant signal-to-distortion ratio (SI-SDR) [46], perceptual evaluation of speech quality (PESQ) [47], extended short-time objective intelligibility (eSTOI) [48] and word error rates (WER) as the evaluation metrics for SMS-WSJ, and report WER for LibriCSS. For PESQ, we report narrow-band MOS-LQO scores based on the ITU P.862.1 standard [44] using the *python-pesq* toolkit.

TABLE I
SI-SDR, PESQ, ESTOI AND WER ON SMS-WSJ TEST SET.

| Approaches | Use $\|Y_q\|$? | SI-SDR (dB) | | | | PESQ | | | | eSTOI | | | | WER (%) | | | |
|---|---|---|---|---|---|---|---|---|---|---|---|---|---|---|---|---|---|
| #mics | - | 1 | 2 | 3 | 6 | 1 | 2 | 3 | 6 | 1 | 2 | 3 | 6 | 1 | 2 | 3 | 6 |
| Unprocessed | - | -5.5 | - | - | - | 1.50 | - | - | - | 0.441 | - | - | - | 78.42 | - | - | - |
| $\text{SISO}_1$ |  | 5.7 | - | - | - | 2.40 | - | - | - | 0.748 | - | - | - | 28.67 | - | - | - |
| $\text{SISO}_1\text{-BF}$ |  | - | -1.4 | 1.5 | 4.8 | - | 1.64 | 1.76 | 1.98 | - | 0.541 | 0.618 | 0.697 | - | 66.71 | 53.41 | 31.42 |
| $\text{SISO}_1\text{-BF-SISO}_2$ |  | - | 8.9 | 10.2 | 11.6 | - | 2.99 | 3.21 | 3.36 | - | 0.848 | 0.879 | 0.898 | - | 14.87 | 12.06 | 10.48 |
| $\text{MISO}_1$ | No | - | 8.2 | 8.9 | 10.2 | - | 2.85 | 2.98 | 3.05 | - | 0.826 | 0.844 | 0.859 | - | 17.19 | 14.99 | 13.95 |
| $\text{MISO}_1\text{-BF}$ |  | - | -1.2 | 1.9 | 5.8 | - | 1.64 | 1.77 | 2.01 | - | 0.543 | 0.624 | 0.716 | - | 66.39 | 53.36 | 29.34 |
| $\text{MISO}_1\text{-BF-MISO}_2$ |  | - | 9.8 | 11.6 | 13.6 | - | 3.13 | 3.20 | 3.51 | - | 0.866 | 0.885 | 0.919 | - | 13.78 | 12.27 | 9.49 |
| $\text{MISO}_1\text{-BF-MISO}_3$ |  | - | **12.7** | **14.0** | 15.2 | - | **3.43** | **3.60** | **3.76** | - | **0.907** | **0.926** | **0.942** | - | **10.67** | 9.66 | **8.24** |
| $\text{MISO}_1+\text{MISO}_4$ |  | - | 9.5 | 10.4 | 11.7 | - | 3.09 | 3.05 | 3.21 | - | 0.860 | 0.864 | 0.885 | - | 14.35 | 14.31 | 12.46 |
| $\text{MISO}_1+\text{MISO}_5$ |  | - | 11.4 | 12.4 | 13.2 | - | 3.33 | 3.41 | 3.53 | - | 0.895 | 0.905 | 0.918 | - | 11.59 | 10.97 | 10.01 |
| $\text{SISO}_1$ |  | 5.9 | - | - | - | 2.44 | - | - | - | 0.753 | - | - | - | 26.86 | - | - | - |
| $\text{SISO}_1\text{-BF}$ |  | - | -1.4 | 1.5 | 4.9 | - | 1.64 | 1.76 | 1.98 | - | 0.541 | 0.618 | 0.698 | - | 66.51 | 53.70 | 31.10 |
| $\text{SISO}_1\text{-BF-SISO}_2$ |  | - | 8.9 | 10.9 | 12.2 | - | 3.02 | 3.27 | 3.44 | - | 0.854 | 0.887 | 0.906 | - | 13.95 | 11.50 | 10.08 |
| $\text{MISO}_1$ | Yes | - | 8.1 | 8.6 | 10.2 | - | 2.84 | 2.92 | 3.06 | - | 0.823 | 0.838 | 0.862 | - | 17.38 | 15.64 | 13.92 |
| $\text{MISO}_1\text{-BF}$ |  | - | -1.2 | 1.8 | 5.9 | - | 1.64 | 1.77 | 2.01 | - | 0.543 | 0.624 | 0.717 | - | 66.61 | 53.41 | 29.04 |
| $\text{MISO}_1\text{-BF-MISO}_2$ |  | - | 10.2 | 11.0 | 13.4 | - | 3.02 | 3.29 | 3.41 | - | 0.859 | 0.889 | 0.910 | - | 14.82 | 11.68 | 9.98 |
| $\text{MISO}_1\text{-BF-MISO}_3$ |  | - | 12.3 | 13.8 | **15.6** | - | 3.39 | 3.59 | **3.76** | - | 0.903 | 0.925 | **0.942** | - | 11.39 | **9.45** | 8.28 |
| $\text{MISO}_1+\text{MISO}_4$ |  | - | 9.7 | 10.3 | 11.6 | - | 2.95 | 3.03 | 3.19 | - | 0.848 | 0.863 | 0.883 | - | 16.17 | 14.21 | 12.74 |
| $\text{MISO}_1+\text{MISO}_5$ |  | - | 11.5 | 12.4 | 13.2 | - | 3.32 | 3.41 | 3.52 | - | 0.893 | 0.905 | 0.918 | - | 12.06 | 10.94 | 9.92 |
| 1ch DPRNN-TasNet [44] |  | 6.5 | - | - | - | 2.28 | - | - | - | 0.734 | - | - | - | 38.12 | - | - | - |
| FaSNet + TAC + joint + 4ms [5] | - | - | 6.9 | 7.6 | 8.6 | - | 2.27 | 2.31 | 2.37 | - | 0.731 | 0.749 | 0.771 | - | 34.84 | 32.31 | 29.8 |
| Multi-channel Conv-TasNet [45] |  | - | 5.8 | 9.0 | 10.8 | - | 2.16 | 2.60 | 2.78 | - | 0.720 | 0.810 | 0.844 | - | 45.72 | 26.10 | 23.05 |
| 6ch spatial clustering (cACGMM) [27] |  | - | - | - | - | - | - | - | - | - | - | - | - | - | - | - | 39.00 |
| 6ch spatial clustering (cACGMM) with MVDR [27] | - | - | - | - | - | - | - | - | - | - | - | - | - | - | - | - | 18.70 |
| Oracle direct sound + early reflections | - | - | - | - | - | - | - | - | - | - | - | - | - | 7.04 | - | - | - |
| Oracle monaural SMM ($\|S_q\|/\|Y_q\|$) [13] | - | 1.8 | - | - | - | 3.37 | - | - | - | 0.904 | - | - | - | 6.74 | - | - | - |
| Oracle monaural PSM ($\|S_q\|\cos(\angle S_q - \angle Y_q)/\|Y_q\|$) [13] | - | 6.0 | - | - | - | 3.65 | - | - | - | 0.902 | - | - | - | 6.51 | - | - | - |
| Oracle direct sound | - | - | - | - | - | - | - | - | - | - | - | - | - | 6.40 | - | - | - |

The time-domain signal corresponding to $S_q(c)$ is used as the references for metric computation.

## VII. EVALUATION RESULTS

This section reports evaluation results on SMS-WSJ and LibriCSS. Based on SMS-WSJ, we analyze the robustness of our trained models to geometry mismatches in Section VII.B.

In our experiments on LibriCSS, we find that including the magnitude features of the reference microphone in addition to the RI components for model training shows better ASR performance. The rationale is that the pattern of magnitude is more stable than RI components (or waveforms), and including it could lead to more robust separation on real-recorded data. We also add magnitude features for our experiments on SMS-WSJ. Although adding these features does not make a big difference in performance as the training and testing sets of SMS-WSJ are matched, we add these results for completeness. More specifically, for example for $\text{MISO}_1$, we use $\langle Y_q, ..., Y_P, Y_1, ..., Y_{q-1}, |Y_q| \rangle$ to predict $\langle S_q(1), ..., S_q(C) \rangle$. For the architecture in Figure 6, adding this feature only introduces $1 \times 24 \times 3 \times 3$ parameters (to the first convolutional layer).

### A. Results on SMS-WSJ

TABLE I reports the performance of single- and multi-channel separation and dereverberation on SMS-WSJ, along with oracle results such as direct sound, direct sound plus early reflections, and oracle T-F masks such as the spectral magnitude mask (SMM) and phase-sensitive mask (PSM) [13].

We mainly comment on the single- and six-channel results, because similar trends are observed in the two- and three-channel cases. We only go over the results obtained by including magnitude features. We observe clear improvement using $\text{MISO}_1$ over $\text{SISO}_1$ (10.2 vs. 5.9 dB SI-SDR), suggesting that MISO is capable of exploiting spatial in addition to spectral information on fixed-geometry arrays. Comparing $\text{MISO}_1\text{-BF}$ and $\text{SISO}_1\text{-BF}$ (5.9 vs. 4.9 dB SI-SDR), we find that $\text{MISO}_1$ produces better covariances for MVDR beamforming. By using MISO for post-filtering, $\text{MISO}_1\text{-BF-MISO}_2$ produces much better performance over $\text{MISO}_1$ and $\text{MISO}_1\text{-BF}$ (13.4 vs. 10.2 and 5.9 dB SI-SDR), and is 1.2 dB better than $\text{SISO}_1\text{-BF-SISO}_2$ (13.4 vs. 12.2 dB), indicating the benefit of replacing the two single-microphone networks in SISO-BF-SISO with MISO. By predicting target speakers one by one, $\text{MISO}_1\text{-BF-MISO}_3$ further improves SI-SDR to 15.6 dB, amounting to 2.2 dB improvement over $\text{MISO}_1\text{-BF-MISO}_2$ (15.6 vs. 13.4 dB), 9.7 dB improvement over single-channel processing (15.6 vs. 5.9 dB), and 21.1 dB improvement over no processing (15.6 vs. -5.5 dB). Similar trends are observed from PESQ, eSTOI and WER results. MISO$_1$-BF-MISO$_3$ yields 8.28% WER, which is very close to the 6.40% WER obtained by using the oracle direct sound of each source for decoding. MISO$_1$-BF-MISO$_2$ produces clearly better performance than MISO$_1$-MISO$_4$ (13.4 vs. 11.6 dB), and MISO$_1$-BF-MISO$_3$ is also clearly better than MISO$_1$-MISO$_5$ (15.6 vs. 13.2 dB). These two comparisons show the effectiveness of using a TI-MVDR module between the two networks.

Our algorithm shows much better WER over conventional spatial clustering based on cACGMM with or without further MVDR beamforming (8.28% vs. 18.7% and 39.0% WER).

Compared with monaural DPRNN-TasNet [44], our SISO$_1$ model shows clearly better PESQ and WER (2.44 vs. 2.28 and 26.86% vs. 38.12%) and slightly better eSTOI (0.753 vs. 0.734), and worse SI-SDR (5.9 vs. 6.5 dB) which is a time-domain metric. Our multi-channel models such as MISO$_1$ and MISO$_1$-BF-MISO$_3$ produce much better performance on all the four metrics over FaSNet with TAC modules [5] (10.2 and 15.6 vs. 8.6 dB in SI-SDR, 3.06 and 3.76 vs. 2.37 in PESQ, 0.862 and 0.942 vs. 0.771 in eSTOI, and 13.92% and 8.28% vs. 29.8% in WER). Multi-channel Conv-TasNet [45] produces strong SI-SDR results in the three- and six-microphone cases, but not PESQ, eSTOI and WER results. Note that, similarly to the proposed algorithms, FaSNet and multi-channel Conv-TasNet also have the advantage of knowing array geometry.

TABLE II compares the performance of MISO$_1$ with the time-varying MVDR beamformer detailed in Section VI.F. In both two- and six-channel cases, the time-varying beamformer (denoted as MISO$_1$-tvBF) shows better performance over MISO$_1$-BF, but is worse than MISO$_1$.

These results indicate the outstanding effectiveness of our proposed algorithms on fixed-geometry arrays. A sound demo page[1] is available online for comparing different systems.

### B. Sensitivity to Geometry Mismatch

To investigate the sensitivity of our trained models to geometry mismatches, we add small perturbations to the microphone positions in SMS-WSJ to simulate manufacturing errors. For each microphone position of each test mixture, the perturbations (in millimeters) along the X, Y and Z axis are independently sampled from a Gaussian distribution with zero mean and standard deviation $\sigma$. Table III reports the results. Such perturbations have no influence on SISO$_1$ and SISO$_1$-BF-SISO$_2$. This is as expected since the models are essentially monaural, although SISO$_1$-BF-SISO$_2$ considers MVDR results as a spatial feature for SISO$_2$. The perturbations only slightly deteriorate the performance of MISO$_1$ and MISO$_1$-BF-MISO$_3$. Even for a large $\sigma$ (5 mm), which is unlikely to happen in modern manufacturing, our models still perform comparably well to not applying any perturbations.

### C. Results on LibriCSS

TABLE IV presents utterance-wise evaluation results on the seven-channel task of LibriCSS. We observe that using magnitude features leads to clear improvements for MISO$_1$.

[1]https://zqwang7.github.io/demos/SMSWSJ_demo/taslp20_SMSWSJ_demo.

TABLE II
SI-SDR, PESQ, eSTOI AND WER COMPARISON OF MISO$_1$ WITH A TIME-VARYING MVDR ON SMS-WSJ TEST SET (INCLUDING MAGNITUDE FEATURES).

| Approaches | $\Delta$ | SI-SDR (dB) | | PESQ | | eSTOI | | WER (%) | |
|---|---|---|---|---|---|---|---|---|---|
| #mics | - | 2 | 6 | 2 | 6 | 2 | 6 | 2 | 6 |
| Unprocessed | - | -5.5 | -5.5 | 1.5 | 1.5 | 0.441 | 0.441 | 78.42 | 78.42 |
| MISO$_1$-BF | - | -1.2 | 5.9 | 1.64 | 2.01 | 0.543 | 0.717 | 66.61 | 29.04 |
| MISO$_1$-tvBF | 0 | 1.4 | 9.6 | 1.79 | 2.87 | 0.619 | 0.836 | 56.01 | 19.63 |
| MISO$_1$-tvBF | 1 | 0.5 | 9.0 | 1.71 | 2.56 | 0.588 | 0.813 | 60.84 | 24.47 |
| MISO$_1$-tvBF | 2 | -0.1 | 8.4 | 1.68 | 2.37 | 0.571 | 0.791 | 62.38 | 27.23 |
| MISO$_1$-tvBF | 3 | -0.3 | 8.0 | 1.67 | 2.26 | 0.564 | 0.776 | 63.02 | 27.91 |
| MISO$_1$ | - | 8.1 | 10.2 | 2.84 | 3.06 | 0.823 | 0.862 | 17.38 | 13.92 |

TABLE III
SI-SDR (dB) ON SMS-WSJ TEST SET WITH PERTURBED MICROPHONE POSITIONS (6CH, INCLUDING MAGNITUDE FEATURES).

| Approaches | $\sigma$ (mm) | | | | | |
|---|---|---|---|---|---|---|
| | 0 | 1 | 2 | 3 | 4 | 5 |
| Unprocessed | -5.5 | -5.5 | -5.5 | -5.5 | -5.5 | -5.5 |
| SISO$_1$ (1ch) | 5.9 | 5.9 | 5.9 | 5.9 | 5.9 | 5.9 |
| SISO$_1$-BF-SISO$_2$ | 12.2 | 12.2 | 12.2 | 12.2 | 12.2 | 12.2 |
| MISO$_1$ | 10.2 | 10.2 | 10.1 | 10.1 | 9.9 | 9.8 |
| MISO$_1$-BF-MISO$_3$ | 15.6 | 15.6 | 15.5 | 15.3 | 15.2 | 15.0 |

Using MISO based post-filtering that predicts target speakers one by one, MISO$_1$-BF-MISO$_3$ yields large improvements over MISO$_1$, especially for high overlap ratios (e.g. 8.3% vs. 13.0% WER on 40% overlap).

Table V reports the performance of monaural processing on the utterance-wise task of LibriCSS. Using magnitude features in SISO$_1$ also leads to some improvement, for example from 10.0% to 9.2% WER in the 0S condition. The SISO$_1$-SISO$_3$ system stacks two SISO networks, where SISO$_1$ resolves the permutation problem and SISO$_3$ predicts target speakers one by one by using $\langle Y_q, \hat{S}_q^{(1)}(c) \rangle$ as inputs to estimate $S_q(c)$. Better performance is observed by using a second SISO network.

TABLE VI and TABLE VII respectively present the continuous evaluation results on the seven- and one-channel tasks of LibriCSS. We observe that using magnitude features in MISO$_1$ and SISO$_1$ also helps. MISO$_1$+SC means that we use a dedicated speaker counting (SC) network to count speakers at each frame, and use the counting results to merge MISO$_1$ outputs. Clear improvement is obtained over MISO$_1$. Similar to the utterance-wise evaluation, MISO$_1$-BF-MISO$_3$+SC, which applies speaker counting results to merge the outputs of MISO$_3$, produces clearly better performance over MISO$_1$+SC.

Compared with monaural models, our seven-channel models yield large improvements in both utterance-wise and continuous-input evaluations. These results clearly demonstrate the effectiveness of DNN and MISO based time-varying non-linear beamforming and post-filtering, and most importantly, the strong generalizability of our trained models to real arrays with the same geometry.

Based on the default ASR backend, our best seven-channel frontend produces much better WER on LibriCSS over the current best results reported in [41], which uses magnitudes and IPDs to compute block-online masking based MVDR for separation. For example, on 40% overlap, MISO$_1$-BF-

TABLE IV
WER (%) on LibriCSS (Utterance-Wise Evaluation, 7ch).

| Approaches | Use $|Y_q|$? | ASR Backend | Overlap Ratio (%) | | | | | |
|---|---|---|---|---|---|---|---|---|
| | | | 0S | 0L | 10 | 20 | 30 | 40 |
| Unprocessed | - | Default | 11.8 | 11.7 | 18.8 | 27.2 | 35.6 | 43.3 |
| MISO$_1$ | No | | 9.3 | 10.3 | 9.6 | 10.9 | 12.6 | 13.9 |
| MISO$_1$ | Yes | | 7.7 | 7.5 | 7.9 | 9.6 | 11.3 | 13.0 |
| MISO$_1$-BF-MISO$_3$ | Yes | | 5.8 | 5.8 | 5.9 | 6.5 | 7.7 | 8.3 |
| MISO$_1$-BF-MISO$_3$ | Yes | Our E2E | 3.4 | 3.5 | **3.2** | **3.9** | **4.7** | **4.7** |
| Chen et al. [21] | - | Default | 8.3 | 8.4 | 11.6 | 16.0 | 18.4 | 21.6 |
| Chen et al. [41] | - | Default | 7.2 | 7.5 | 9.6 | 11.3 | 13.7 | 15.1 |
| Chen et al. [41] | - | E2E [41] | **3.1** | **3.3** | 3.7 | 4.8 | 5.6 | 6.2 |
| Oracle anechoic speech | - | Default | 4.9 | 5.1 | - | - | - | - |

TABLE V
WER (%) on LibriCSS (Utterance-Wise Evaluation, 1ch).

| Approaches | Use $|Y_q|$? | ASR Backend | Overlap Ratio (%) | | | | | |
|---|---|---|---|---|---|---|---|---|
| | | | 0S | 0L | 10 | 20 | 30 | 40 |
| Unprocessed | - | Default | 11.8 | 11.7 | 18.8 | 27.2 | 35.6 | 43.3 |
| SISO$_1$ | No | | 10.0 | 9.7 | 11.8 | 16.7 | 20.5 | 24.1 |
| SISO$_1$ | Yes | | 9.2 | 8.9 | 11.6 | 15.5 | 20.0 | 23.1 |
| SISO$_1$-SISO$_3$ | Yes | | 9.1 | 8.6 | 10.6 | 13.9 | 17.1 | 19.8 |
| SISO$_1$-SISO$_3$ | Yes | Our E2E | **5.4** | 5.3 | **6.5** | **8.6** | **11.2** | **12.3** |
| Chen et al., [21] | - | Default | 12.7 | 12.1 | 17.6 | 23.2 | 30.5 | 35.6 |
| Chen et al., [41] | - | Default | 12.9 | 12.2 | 15.1 | 20.1 | 24.3 | 27.6 |
| Chen et al., [41] | - | E2E [41] | **5.4** | **5.0** | 7.5 | 10.7 | 13.8 | 17.1 |

TABLE VI
WER (%) on LibriCSS (Continuous-Input Evaluation, 7ch).

| Approaches | Use $|Y_q|$? | ASR Backend | Overlap Ratio (%) | | | | | |
|---|---|---|---|---|---|---|---|---|
| | | | 0S | 0L | 10 | 20 | 30 | 40 |
| Unprocessed | - | Default | 15.4 | 11.5 | 21.7 | 27.0 | 34.3 | 40.5 |
| MISO$_1$ | No | | 13.1 | 14.0 | 13.2 | 13.7 | 16.4 | 17.1 |
| MISO$_1$+SC | No | | 7.5 | 8.9 | 8.6 | 10.9 | 12.8 | 14.9 |
| MISO$_1$ | Yes | Default | 10.7 | 10.5 | 10.9 | 11.5 | 13.8 | 15.3 |
| MISO$_1$+SC | Yes | | 7.9 | 8.5 | 8.5 | 10.5 | 12.3 | 14.3 |
| MISO$_1$-BF-MISO$_3$+SC | Yes | | 7.7 | 7.5 | 7.4 | 8.4 | 9.7 | 11.3 |
| MISO$_1$-BF-MISO$_3$+SC | Yes | Our E2E | 5.4 | 5.0 | **4.8** | **5.0** | **6.6** | **7.6** |
| Chen et al. [21] | - | Default | 11.9 | 9.7 | 13.4 | 15.1 | 19.7 | 22.0 |
| Chen et al. [41] | - | Default | 11.0 | 8.7 | 12.6 | 13.5 | 17.6 | 19.6 |
| Chen et al., [41] | - | E2E [41] | **5.2** | **4.0** | 5.8 | 6.8 | 9.0 | 10.0 |

TABLE VII
WER (%) on LibriCSS (Continuous-Input Evaluation, 1ch).

| Approaches | Use $|Y_q|$? | ASR Backend | Overlap Ratio (%) | | | | | |
|---|---|---|---|---|---|---|---|---|
| | | | 0S | 0L | 10 | 20 | 30 | 40 |
| Unprocessed | - | Default | 15.4 | 11.5 | 21.7 | 27.0 | 34.3 | 40.5 |
| SISO$_1$ | No | | 12.3 | 12.2 | 13.6 | 16.8 | 21.1 | 23.8 |
| SISO$_1$+SC | No | | 9.9 | 11.2 | 11.8 | 15.6 | 19.7 | 23.1 |
| SISO$_1$ | Yes | Default | 12.2 | 12.1 | 13.2 | 16.4 | 20.6 | 23.2 |
| SISO$_1$+SC | Yes | | 9.4 | 9.7 | 11.6 | 15.2 | 19.7 | 23.0 |
| SISO$_1$-SISO$_3$+SC | Yes | | 10.7 | 10.4 | 11.7 | 14.8 | 18.8 | 20.8 |
| SISO$_1$-SISO$_3$+SC | Yes | Our E2E | 7.1 | **6.1** | **7.5** | **9.5** | **12.7** | **15.2** |
| Chen et al., [21] | - | Default | 17.6 | 16.3 | 20.9 | 26.1 | 32.6 | 36.1 |
| Chen et al., [41] | - | Default | 13.3 | 11.7 | 16.3 | 20.7 | 25.6 | 29.3 |
| Chen et al., [41] | - | E2E [41] | **6.9** | **6.1** | 9.1 | 12.5 | 16.7 | 19.3 |

MISO$_3$+SC obtains 11.3% WER in continous-input evaluation and MISO$_1$-BF-MISO$_3$ gets 8.3% in utterance-wise evaluation. They are much better than 19.6% and 15.1% WER reported in [41]. We believe that TI-MVDR alone cannot sufficiently suppress non-target speakers, although it maintains each target speaker distortionlessly. Our single-channel models, which combine a TCN with a dense U-Net for complex spectral mapping, also obtain better performance over the monaural ones in [21] and [41], which use BLSTM and conformer for real-valued T-F masking.

We further apply an end-to-end (E2E) ASR backend for ASR. It is a conformer-based model trained on the LibriSpeech corpus using ESPnet. On the *test-clean* set of LibriSpeech, the model obtains 2.1% WER, which is almost the same as the 2.08% WER obtained by the E2E ASR backend used in [41]. Combined with our frontend, the E2E backend gets overall better results than [41] (for example on 40% overlap, 4.7% vs. 6.2% WER in Table IV, 12.3% vs. 17.1% in Table V, 7.6% vs. 10.0% in TABLE VI, and 15.2% vs. 19.3% in TABLE VII), although the performance is slightly worse (or comparable) in the 0S and 0L conditions.

## VIII. CONCLUDING REMARKS

We have proposed a multi-microphone complex spectral mapping approach to address both speaker separation and dereverberation. The superior separation and ASR results on SMS-WSJ indicate that two-speaker separation in simulated reverberant conditions can now be addressed very well by exploiting the spectral and spatial information afforded by a fixed six-microphone array using frequency-domain methods, even though our study considers dereverberation in addition to separation and does not leverage extra information such as speaker embeddings and visual cues.

Time-domain approaches recently gain popularity in monaural speaker separation. On SMS-WSJ, our single-channel models show better PESQ, eSTOI and WER, and worse SI-SDR over monaural DP-RNN. Our multi-channel models obtain much better PESQ, eSTOI and WER results and competitive SI-SDR results compared with FaSNet with TAC modules, and multi-channel Conv-TasNet. Similarly to the proposed methods, FaSNet and multi-channel Conv-TasNet also have the advantage of knowing array geometry. Importantly, our models also produce strong recognition performance on the more realistic LibriCSS corpus.

Although trained on simulated RIRs, the proposed MISO and MISO-BF-MISO models generalize well to the real device used in LibriCSS. This is a significant finding, as it suggests that we can train models on simulated multi-channel conditions, which can be readily simulated, and expect them to generalize well to real devices with matched array geometry.

Our study shows that using the direct outputs from a strong DNN can produce much better ASR results over time-invariant beamforming, at least in speaker separation in reverberant conditions with weak and relatively stationary noise. This finding contrasts that in single-speaker robust ASR [16], [26], where only one speaker is assumed active in noisy-reverberant environments. The reason could be that multi-talker speech is more harmful for recognition, and therefore frontend processing needs to dramatically suppress non-target speakers. In addition, competing speakers are easier to suppress as speech signal shows strong patterns unlike reverberation and noise. On the other hand, for single-speaker robust ASR, there is only one active speaker and speech distortion is more of a concern, as multi-condition training can deal with noise and reverberation to some extent. Future research shall consider multi-speaker ASR in reverberant conditions with challenging noises. In addition, the application of this approach to multi-channel speech enhancement is straightforward.

The major limitation of our current study for CSS comes from the assumption that each short processing block contains at most two concurrent speakers. To deal with more than two speakers, we could just do say 3- or 4-speaker PIT, or use recursive separation [49] in each block. Another weakness is that the first MISO network needs to run $P$ times, once for each microphone to compute the statistics for beamforming in order to get the best performance, resulting in high computational costs. One solution would be to replace it with a MIMO network that can predict all the target speakers at all the microphones [28]. Another possible way is to run MISO only for the reference microphone and use mask-based beamforming [31], at a cost of some performance degradation.

## REFERENCES

[1] J. R. Hershey, Z. Chen, J. Le Roux, and S. Watanabe, "Deep Clustering: Discriminative Embeddings for Segmentation and Separation," in *IEEE International Conference on Acoustics, Speech and Signal Processing*, 2016, pp. 31–35.

[2] M. Kolbæk, D. Yu, Z.-H. Tan, and J. Jensen, "Multi-Talker Speech Separation with Utterance-Level Permutation Invariant Training of Deep Recurrent Neural Networks," *IEEE/ACM Trans. Audio, Speech, Lang. Process.*, vol. 25, no. 10, pp. 1901–1913, 2017.

[3] Z.-Q. Wang and D. L. Wang, "Combining Spectral and Spatial Features for Deep Learning Based Blind Speaker Separation," *IEEE/ACM Trans. Audio, Speech, Lang. Process.*, vol. 27, no. 2, pp. 457–468, 2019.

[4] T. Yoshioka, H. Erdogan, Z. Chen, and F. Alleva, "Multi-Microphone Neural Speech Separation for Far-Field Multi-Talker Speech Recognition," in *IEEE International Conference on Acoustics, Speech and Signal Processing*, 2018, pp. 5739–5743.

[5] Y. Luo, Z. Chen, N. Mesgarani, and T. Yoshioka, "End-to-End Microphone Permutation and Number Invariant Multi-Channel Speech Separation," in *IEEE International Conference on Acoustics, Speech and Signal Processing*, 2020, pp. 6394–6398.

[6] R. Gu et al., "Enhancing End-to-End Multi-Channel Speech Separation Via Spatial Feature Learning," in *IEEE International Conference on Acoustics, Speech and Signal Processing*, 2020, pp. 7319–7323.

[7] Z.-Q. Wang, K. Tan, and D. Wang, "Deep Learning Based Phase Reconstruction for Speaker Separation: A Trigonometric Perspective," in *IEEE International Conference on Acoustics, Speech and Signal Processing*, 2019, pp. 71–75.

[8] Y. Luo and N. Mesgarani, "Conv-TasNet: Surpassing Ideal Time-Frequency Magnitude Masking for Speech Separation," *IEEE/ACM Trans. Audio, Speech, Lang. Process.*, vol. 27, no. 8, pp. 1256–1266, 2019.

[9] Y. Liu and D. L. Wang, "Divide and Conquer: A Deep CASA Approach to Talker-Independent Monaural Speaker Separation," *IEEE/ACM Trans. Audio, Speech, Lang. Process.*, vol. 27, pp. 2092–2102, 2019.

[10] N. Zeghidour and D. Grangier, "Wavesplit: End-to-End Speech Separation by Speaker Clustering," in *arXiv preprint arXiv:2002.08933*, 2020.

[11] A. Ephrat et al., "Looking to Listen at the Cocktail Party: A Speaker-Independent Audio-Visual Model for Speech Separation," *ACM Trans. Graph.*, vol. 37, no. 4, Apr. 2018.

[12] S. Gannot, E. Vincent, S. Markovich-Golan, and A. Ozerov, "A Consolidated Perspective on Multi-Microphone Speech Enhancement and Source Separation," *IEEE/ACM Trans. Audio, Speech, Lang. Process.*, vol. 25, pp. 692–730, 2017.

[13] D. Wang and J. Chen, "Supervised Speech Separation Based on Deep Learning: An Overview," *IEEE/ACM Trans. Audio, Speech, Lang. Process.*, vol. 26, pp. 1702–1726, 2018.

[14] J. Heymann, L. Drude, A. Chinaev, and R. Haeb-Umbach, "BLSTM Supported GEV Beamformer Front-End for The 3rd CHiME Challenge," in *IEEE Workshop on Automatic Speech Recognition and Understanding*, 2015, pp. 444–451.

[15] T. Yoshioka et al., "The NTT CHiME-3 System: Advances in Speech Enhancement and Recognition for Mobile Multi-Microphone Devices," in *IEEE Workshop on Automatic Speech Recognition and Understanding*, 2015, pp. 436–443.

[16] J. Barker, R. Marxer, E. Vincent, and S. Watanabe, "The Third 'CHiME' Speech Separation and Recognition Challenge: Analysis and Outcomes," *Comput. Speech Lang.*, vol. 46, pp. 605–626, 2017.

[17] Z.-Q. Wang, X. Zhang, and D. L. Wang, "Robust Speaker Localization Guided by Deep Learning Based Time-Frequency Masking," *IEEE/ACM Trans. Audio, Speech, Lang. Process.*, vol. 27, no. 1, pp. 178–188, 2019.

[18] Z.-Q. Wang, J. Le Roux, and J. R. Hershey, "Multi-Channel Deep Clustering: Discriminative Spectral and Spatial Embeddings for Speaker-Independent Speech Separation," in *IEEE International Conference on Acoustics, Speech and Signal Processing*, 2018, pp. 1–5.

[19] S. Chakrabarty and E. A. P. Habets, "Time-Frequency Masking Based Online Multi-Channel Speech Enhancement with Convolutional Recurrent Neural Networks," *IEEE J. Sel. Top. Signal Process.*, 2019.

[20] S. Chakrabarty and E. A. P. Habets, "Multi-Speaker DOA Estimation using Deep Convolutional Networks Trained with Noise Signals," *IEEE J. Sel. Top. Signal Process.*, vol. 13, no. 1, pp. 8–21, 2019.

[21] Z. Chen et al., "Continuous Speech Separation: Dataset and Analysis," in *IEEE International Conference on Acoustics, Speech and Signal Processing*, 2020, pp. 7284–7288.

[22] Z.-Q. Wang and D. Wang, "Deep Learning Based Target Cancellation for Speech Dereverberation," *IEEE/ACM Trans. Audio, Speech, Lang. Process.*, vol. 28, pp. 941–950, 2020.

[23] Z.-Q. Wang, P. Wang, and D. Wang, "Complex Spectral Mapping for Single- and Multi-Channel Speech Enhancement and Robust ASR," *IEEE/ACM Trans. Audio, Speech, Lang. Process.*, vol. 28, pp. 1778–1787, 2020.

[24] Y. Kubo, T. Nakatani, M. Delcroix, K. Kinoshita, and S. Araki, "Mask-Based MVDR Beamformer for Noisy Multisource Environments: Introduction of Time-Varying Spatial Covariance Model," in *IEEE International Conference on Acoustics, Speech and Signal Processing*, 2019, pp. 6855–6859.

[25] Y. Luo, C. Han, N. Mesgarani, E. Ceolini, and S.-C. Liu, "FaSNet: Low-Latency Adaptive Beamforming for Multi-Microphone Audio Processing," in *IEEE Workshop on Automatic Speech Recognition and Understanding*, 2019, pp. 260–267.

[26] R. Haeb-Umbach, J. Heymann, L. Drude, S. Watanabe, M. Delcroix, and T. Nakatani, "Far-Field Automatic Speech Recognition," *Proc. IEEE*, 2020.

[27] L. Drude, J. Heitkaemper, C. Boeddeker, and R. Haeb-Umbach, "SMS-WSJ: Database, Performance Measures, and Baseline Recipe for Multi-Channel Source Separation and Recognition," in *arXiv preprint arXiv:1910.13934*, 2019.

[28] Z.-Q. Wang and D. Wang, "Multi-Microphone Complex Spectral Mapping for Speech Dereverberation," in *IEEE International Conference on Acoustics, Speech and Signal Processing*, 2020, pp. 486–490.

[29] D. S. Williamson, Y. Wang, and D. L. Wang, "Complex Ratio Masking for Monaural Speech Separation," *IEEE/ACM Trans. Audio, Speech, Lang. Process.*, pp. 483–492, 2016.

[30] T. Ochiai, M. Delcroix, R. Ikeshita, K. Kinoshita, T. Nakatani, and S. Araki, "Beam-TasNet: Time-Domain Audio Separation Network Meets Frequency-Domain Beamformer," in *IEEE International Conference on Acoustics, Speech and Signal Processing*, 2020, pp. 6384–6388.

[31] Z.-Q. Wang et al., "Sequential Multi-Frame Neural Beamforming for Speech Separation and Enhancement," in *IEEE Spoken Language Technology Workshop*, 2021, pp. 905–911.

[32] T. Nakatani, C. Boeddeker, K. Kinoshita, R. Ikeshita, M. Delcroix, and R. Haeb-Umbach, "Jointly Optimal Denoising, Dereverberation, and Source Separation," *IEEE/ACM Trans. Audio Speech Lang. Process.*, vol. 28, pp. 2267–2282, 2020.

[33] X. Xiao et al., "Deep Beamforming Networks for Multi-Channel Speech Recognition," in *IEEE International Conference on Acoustics, Speech and Signal Processing*, 2016, pp. 5745–5749.

[34] Z. Zhang, Y. Xu, M. Yu, S.-X. Zhang, L. Chen, and D. Yu, "ADL-MVDR: All Deep Learning MVDR Beamformer for Target Speech Separation," in *arXiv preprint arXiv:2008.06994*, 2020.

[35] D. Stoller, S. Ewert, and S. Dixon, "Wave-U-Net: A Multi-Scale Neural Network for End-to-End Audio Source Separation," in *Proceedings of*

*ISMIR*, 2018, pp. 334–340.
[36] C. Le Liu, S.-W. Fu, Y.-J. Li, J.-W. Huang, H.-M. Wang, and Y. Tsao, "Multichannel Speech Enhancement by Raw Waveform-Mapping using Fully Convolutional Networks," *IEEE/ACM Trans. Audio Speech Lang. Process.*, vol. 28, pp. 1888–1900, 2020.
[37] J. Heitkaemper, D. Jakobeit, C. Boeddeker, L. Drude, and R. Haeb-Umbach, "Demystifying TasNet: A Dissecting Approach," in *IEEE International Conference on Acoustics, Speech and Signal Processing*, 2020, pp. 6359–6363.
[38] S. Bai, J. Z. Kolter, and V. Koltun, "An Empirical Evaluation of Generic Convolutional and Recurrent Networks for Sequence Modeling," in *arXiv preprint arXiv:1803.01271*, 2018.
[39] O. Ronneberger, P. Fischer, and T. Brox, "U-Net: Convolutional Networks for Biomedical Image Segmentation," in *Proceedings of MICCAI*, 2015.
[40] G. Huang, Z. Liu, L. Van Der Maaten, and K. Q. Weinberger, "Densely Connected Convolutional Networks," in *IEEE Conference on Computer Vision and Pattern Recognition*, 2017, vol. 2017-Janua, pp. 2261–2269.
[41] S. Chen *et al.*, "Continuous Speech Separation with Conformer," in *arXiv preprint arXiv:2008.05773v1*, 2020.
[42] E. Habets, "Room Impulse Response Generator," 2010.
[43] V. Manohar, "https://kaldi-asr.org/models/m4." 2018.
[44] Y. Luo, Z. Chen, and T. Yoshioka, "Dual-Path RNN: Efficient Long Sequence Modeling for Time-Domain Single-Channel Speech Separation," in *IEEE International Conference on Acoustics, Speech and Signal Processing*, 2020, pp. 46–50.
[45] J. Zhang, C. Zorila, R. Doddipatla, and J. Barker, "On End-to-End Multi-Channel Time Domain Speech Separation in Reverberant Environments," in *IEEE International Conference on Acoustics, Speech and Signal Processing*, 2020, pp. 6389–6393.
[46] J. Le Roux, S. Wisdom, H. Erdogan, and J. R. Hershey, "SDR – Half-Baked or Well Done?," in *IEEE International Conference on Acoustics, Speech and Signal Processing*, 2019, pp. 626–630.
[47] A. W. Rix, J. G. Beerends, M. P. Hollier, and A. P. Hekstra, "Perceptual Evaluation of Speech Quality (PESQ)-A New Method for Speech Quality Assessment of Telephone Networks and Codecs," in *IEEE International Conference on Acoustics, Speech and Signal Processing*, 2001, vol. 2, pp. 749–752.
[48] J. Jensen and C. H. Taal, "An Algorithm for Predicting The Intelligibility of Speech Masked by Modulated Noise Maskers," *IEEE/ACM Trans. Audio, Speech, Lang. Process.*, vol. 24, no. 11, pp. 2009–2022, 2016.
[49] T. von Neumann *et al.*, "Multi-Talker ASR for An Unknown Number of Sources: Joint Training of Source Counting, Separation and ASR," in *Proceedings of Interspeech*, 2020.

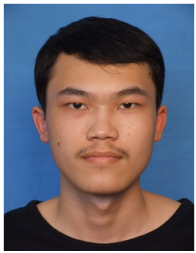
**Zhong-Qiu Wang** (S'16) is a visiting research scientist with Mitsubishi Electric Research Laboratories, Cambridge, USA. He received the B.E. degree in computer science and technology from Harbin Institute of Technology, Harbin, China, in 2013, and the M.S degree and the Ph.D. degree in computer science and engineering from The Ohio State University, Columbus, USA, in 2017 and 2020. His research interests include microphone array processing, speech separation, robust automatic speech recognition, machine learning, and deep learning

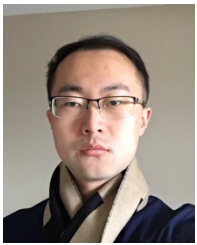
**Peidong Wang** is a senior applied scientist in Microsoft cloud & AI. He received his Ph.D. degree in computer science and engineering from The Ohio State University in 2021, and his B.E. degree in electronic information engineering from University of Science and Technology of China in 2015. His research interests include robust, end-to-end, and efficient automatic speech recognition, and speech separation. He has been serving as a reviewer in various conferences and journals in speech community.

**DeLiang Wang**, photograph and biography not provided at the time of publication.